\documentclass[twocolumn]{autart}    

\usepackage{amsmath,amssymb,amsfonts}
\usepackage{amsmath, bm}

\usepackage{algorithm}
\usepackage[noEnd=true,indLines=true]{algpseudocodex}

\usepackage{graphicx}
\usepackage{xcolor}

\usepackage[nolist]{acronym}
\newacro{ddp}[DDP]{Differential Dynamic Programming}
\newacro{sqp}[SQP]{Sequential Quadratic Programming}
\newacro{lqr}[LQR]{Linear-Quadratic Regulator}
\newacro{mpc}[MPC]{Model Predictive Control}
\newacro{uavs}[UAVs]{Unmanned Aerial Vehicles}
\newacro{sqp}[SQP]{Sequential Quadratic Programming}
\newacro{}[]{}
\newacro{}[]{}
\newacro{}[]{}
\newacro{}[]{}
\newacro{}[]{}
\newacro{}[]{}
\newacro{}[]{}
\newacro{}[]{}
\newacro{}[]{}
\def\RR{{\mathbb R}}

\def\gg{{\mathfrak g}}

\def\se{{\mathfrak s \mathfrak e}}

\def\Gg{{\mathcal G}}
\def\Xx{{\mathcal X}}
\def\Ll{{\mathcal L}}

\def\Pp{{\mathcal P}}

\DeclareMathOperator{\Ad}{Ad}
\DeclareMathOperator{\ad}{ad}
\DeclareMathOperator{\annd}{and}
\DeclareMathOperator{\Exp}{Exp}
\DeclareMathOperator{\expm}{expm}
\DeclareMathOperator{\Log}{Log}
\DeclareMathOperator{\logg}{log}
\DeclareMathOperator{\logm}{logm}

\def\inv{{^{-1}}}
\def\Tra{^{\top}}
\def\Next{^{\prime}}
\graphicspath{{figs/}}

\begin{document}

\begin{frontmatter}

\title{Constrained Trajectory Optimization on Matrix Lie Groups via Lie-Algebraic Differential Dynamic Programming} 

\thanks[footnoteinfo]{This work was supported by the Academy of Finland under Grants 345661 and 347199.}

\author[TAU]{Gokhan Alcan}\ead{gokhan.alcan@tuni.fi},    
\author[NYU]{Fares J. Abu-Dakka}\ead{fa2656@nyu.edu},               
\author[Aalto]{Ville Kyrki}\ead{ville.kyrki@aalto.fi}  

\address[TAU]{Automation Technology and Mechanical Engineering, Tampere University, Finland}  
\address[NYU]{Mechanical Engineering Program, Division of Engineering, New York University Abu Dhabi, United Arab Emirates}
\address[Aalto]{Department of Electrical Engineering and Automation, Aalto University, Finland}


\begin{abstract}                          
Matrix Lie groups are an important class of manifolds commonly used in control and robotics, and optimizing control policies on these manifolds is a fundamental problem. 
In this work, we propose a novel computationally efficient approach for trajectory optimization on matrix Lie groups using an augmented Lagrangian-based constrained discrete Differential Dynamic Programming (DDP). 
The method involves lifting the optimization problem to the Lie algebra during the backward pass and retracting back to the manifold during the forward pass. 
Unlike previous approaches that addressed constraint handling only for specific classes of matrix Lie groups, the proposed method provides a general solution for nonlinear constraint handling across generic matrix Lie groups. 
We evaluate the effectiveness of the proposed DDP method in handling constraints within a mechanical system characterized by rigid body dynamics in SE(3), assessing its computational efficiency compared to existing direct optimization solvers. 
Additionally, the method demonstrates robustness under external disturbances when applied as a Lie-algebraic feedback control policy on SE(3), and in optimizing a quadrotor's trajectory in a challenging realistic scenario. 
Experiments show that the proposed approach effectively manages general constraints defined on configuration, velocity, and inputs during optimization, while also maintaining stability under external disturbances when executing the resultant control policy in closed-loop.
\end{abstract}

\end{frontmatter}

\section{Introduction}
\label{sec:introduction}
Many physical systems such as robots have non-Euclidean configuration spaces. Using local coordinates to model these spaces can result in singularities and degeneracies, where certain system configurations become impossible to represent. For example, using Euler angles to represent rotation can result in a phenomenon known as gimbal lock. In this case, two of the three rotational axes align, leading to a loss of one degree of freedom, which makes tracking of certain movements or transitions between system configurations impossible \cite{Shuster1993,Hemingway2018}. Such singularities can make it difficult to accurately represent transitions between certain system configurations and may require additional techniques to avoid or mitigate these issues.

The geometry of configuration spaces can be effectively modeled using matrix Lie groups, which provide a well-structured and continuous framework for comprehending the structure and motion of the underlying systems \cite{Bloch2015,Lynch2017}. In recent years, geometric control techniques have gained considerable traction in robotics and control systems, as they seamlessly blend differential geometry and control theory \cite{Bullo1999,Bullo2019}. However, current approaches to solving geometric optimal control problems suffer from either reliance on specific problem details to simplify optimality conditions \cite{Bloch2009,Kobilarov2011} or from rapidly increasing computational complexity with an extended temporal horizon \cite{Kobilarov2008,Patil2015}. To overcome these limitations, \ac{ddp} has become a promising approach for numerically solving such problems. However, existing \ac{ddp}-based geometric control techniques have certain limitations, including explicit consideration of only specific matrix Lie groups \cite{Kobilarov2014}, omission of state constraints \cite{Boutselis2020,Teng2022a,Teng2022b}, or handling of constraints limited to particular groups \cite{Liu2022}. 

\begin{figure}[t!]
\vspace{0.2cm}
	\centering
	\def\svgwidth{1\linewidth}
	{
 \fontsize{9}{9}
		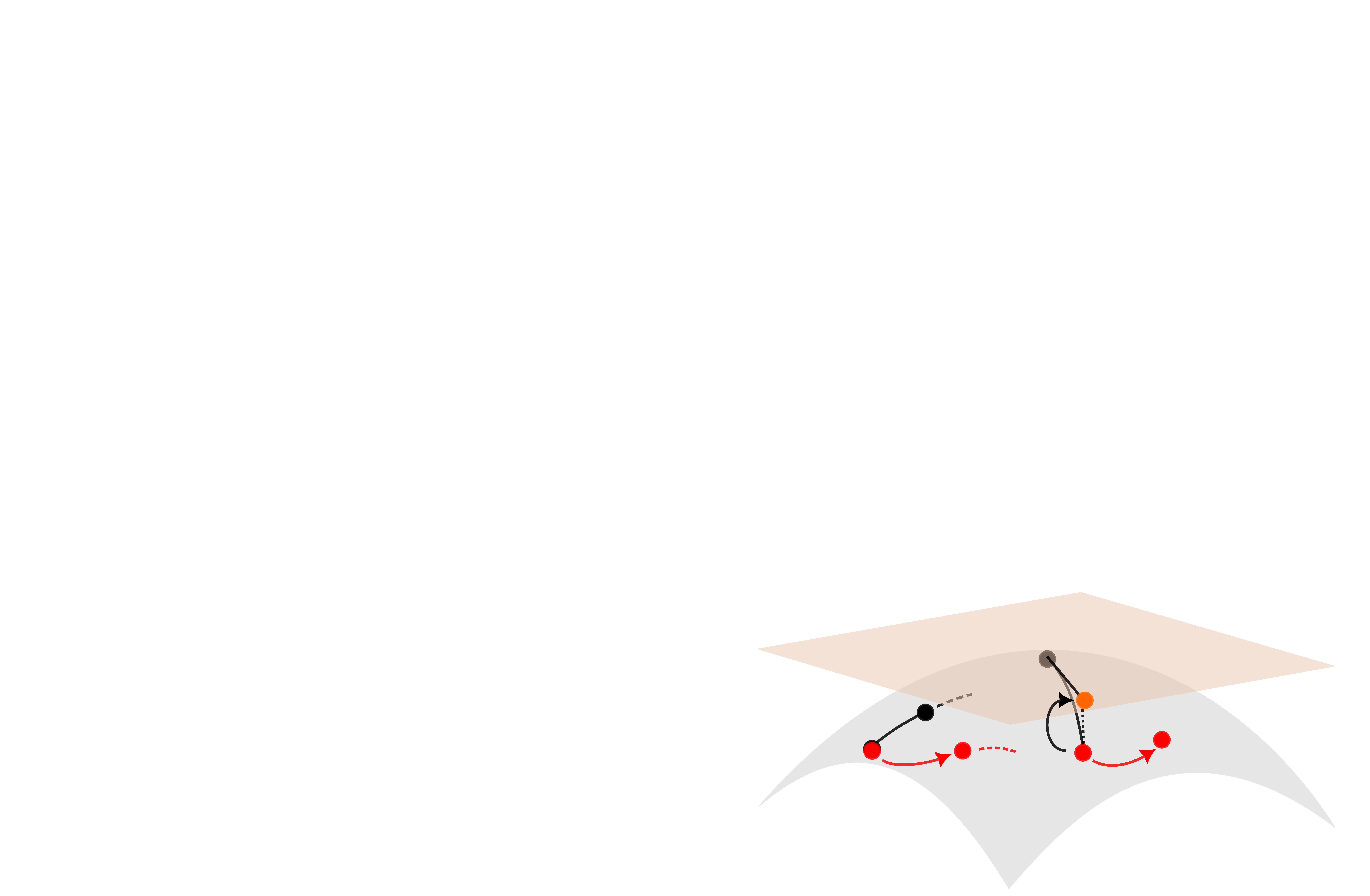}
	\caption{Illustration of the proposed method. a) A trajectory on the configuration manifold is generated using a given nominal input sequence. b) The derivatives of the Q-function are calculated within local tangent spaces. c) The effects of configuration and velocity constraints are also included within local tangent space in the backward pass. d) The forward pass employs closed-loop control to update control inputs. The updated trajectory is then used as the new nominal trajectory for the next iteration.}
	\label{fig:proposed-method}
\end{figure}

In this paper, we present a \ac{ddp}-based geometric control method for matrix Lie groups that incorporates generic nonlinear constraints (Fig. \ref{fig:proposed-method}) overcoming the aforementioned limitations. The main contributions of the work are:
\begin{enumerate}
    \item A novel augmented Lagrangian-based constrained \ac{ddp} algorithm for trajectory optimization on matrix Lie groups.
    \item A principled approach for nonlinear constraint handling for generic matrix Lie groups. This approach overcomes the limitations found in \cite{Liu2022}, which only addressed constraint handling for SO(3) constraints.
    \item Evaluation of the proposed method in handling constraints in a simple mechanical system characterized by rigid body dynamics in SE(3), its computational efficiency compared to existing optimization solvers, and its performance in a realistic quadrotor scenario.
\end{enumerate}

The rest of this paper is organized as follows. Section \ref{sec:related-work} discusses related works from classical control to constrained optimal control employed in smooth manifold systems. In Section \ref{sec:preliminaries}, we provide preliminaries regarding matrix Lie groups. Section \ref{problem-definition} defines the trajectory optimization problem for matrix Lie groups. We detail our proposed method in Section \ref{sec:proposed-method}. In Section \ref{sec:Experiments}, we provide numerical simulation experiments for simple SE(3) and realistic quadrotor dynamics to demonstrate the effectiveness of the approach. Finally, the paper is concluded with potential directions for future work in Section \ref{sec:conclusion}.

\section{Related Work}
\label{sec:related-work}

In recent years, there has been a significant effort to incorporate modern control theory into smooth manifold systems, resulting in a diverse range of theoretical outcomes including stability analysis, controllability, and feedback control design \cite{Bullo1999,Bullo2019}. As an early example, a classical PID-type controller has been successfully applied to fully actuated mechanical systems \cite{Bullo1999}, utilizing configuration and velocity errors defined on a Riemannian manifold. More recently, researchers have developed locally exponentially stable geometric tracking controllers for \ac{uavs} by taking into account the dynamics of rigid bodies on SO(3) \cite{Lee2011} and SE(3) \cite{Lee2010}. Although these methods excel at point stabilization, they do not generate optimal trajectories and require predefined reference trajectories to achieve complex tasks.

Optimization-based control utilizing geometric methods aims to generate optimal trajectories on Riemannian manifolds. Bloch~\etal \cite{Bloch2009} investigated a variational optimal control problem for a 3D rigid body representing dynamics on the Lie group SO(3), while Kobilarov and Marsden \cite{Kobilarov2011} developed necessary conditions for optimal trajectories that correspond to discrete geodesics of a higher-order system and developed numerical methods for their computation. Teng~\etal \cite{Teng2022a} developed a geometric error-state \ac{mpc} for tracking control of systems evolving on connected matrix Lie groups. Even though these methods consider dynamic and kinematic constraints, they lack the ability to account for generic state constraints such as obstacle/configuration avoidance. Saccon~\etal \cite{Saccon2013} derived the projection operator framework on Lie groups, applying it to continuous-time trajectory-optimization problems and utilizing barrier functions for collision avoidance in motion planning for multiple vehicles on SE(3) \cite{Saccon2012}.

In the trajectory optimization field, various studies have aimed to handle constraints in different ways. One of the most straightforward approaches is to formulate a nonlinear program that minimizes the cost function while satisfying all the boundary conditions and constraints. Lee~\etal~\cite{Lee2017} proposed a nonlinear \ac{mpc} framework for spacecraft attitude control on the SO(3) group, where they considered control and exclusion zone constraints. Similarly, Hong~\etal~\cite{Hong2020} combined the Gauss-Newton method with a proximal operator to handle inequality constraints for trajectory optimization on SO(3). Some researchers have focused on optimizing trajectories on other matrix Lie groups, including Kalabic~\etal~\cite{Kalabic2016}, who designed \ac{mpc} on SE(3) by considering only boundary constraints for states and inputs, and Ding~\etal~\cite{Ding2021}, who proposed a representation-free \ac{mpc} for the SE(3) group. However, these methods have limitations, such as being restricted to specific matrix Lie groups and increasing complexity with the number of time instances. Recently, Watterson~\etal~\cite{Watterson2020} proposed a graph search-based safe corridor on manifolds method to optimize trajectories on a Riemannian manifold with obstacles. Although successful in some robotic applications, the method's sensitivity to parameterization affects optimality performance and computational cost. 

Different from the aforementioned \ac{mpc} based direct methods, \ac{ddp} is an indirect numerical method for solving optimal control problems through optimizing the control inputs. Originally proposed by Mayne and Jacobson \cite{Jacobson1970}, \ac{ddp} has been applied to a wide range of complex, high-dimensional systems \cite{Alcan2022}. One of the key advantages of \ac{ddp} is its scalability, which allows it to handle large and complex systems with many degrees of freedom. In addition, \ac{ddp} has a fast convergence rate, which allows it to quickly find near-optimal solutions to the control problem. Another important attribute of \ac{ddp} is its ability to generate feedback control policies, which can be used to implement the optimal control solution in real-time. 

Several early works have investigated the use of \ac{ddp} for geometric control \cite{Kobilarov2014,Kobilarov2015}. These studies focused on specific matrix Lie groups, including SE(3), in order to derive the final form of the \ac{ddp} algorithm for applications in geometric control. Boutselis and Theodorou \cite{Boutselis2020} extended the original \ac{ddp} method by using quadratic expansion schemes for cost functions and dynamics defined on Lie groups. They demonstrated that \ac{ddp} has significantly better convergence rates compared to \ac{sqp} methods. Teng~\etal \cite{Teng2022b} further improved the convergence performance of \ac{ddp} for matrix groups by designing the control objective in its Lie algebra. Both of these approaches \cite{Boutselis2020,Teng2022b} formulate the trajectory optimization on matrix Lie groups in an unconstrained framework. In order to address this limitation, Liu~\etal \cite{Liu2022} extended the work \cite{Boutselis2020} by imposing SO(3) constraints. However, this method is not generalizable to nonlinear constraints for generic matrix Lie groups. This paper aims to solve the problem of handling generic constraints by extending the idea of Lie algebric cost definition \cite{Teng2022b} and developing a \ac{ddp} method for matrix Lie groups under nonlinear constraints.

\section{Preliminaries}
\label{sec:preliminaries}
Consider $\Gg$ is an $n$-dimensional matrix Lie group, and $\gg$ its associated Lie algebra, i.e., its tangent space at the identity.  An isomorphism between the vector space $\RR^n$ and $\gg$ can be defined through the following operators:
\begin{equation}
    \begin{aligned}
        (\cdot)^\wedge & :\RR^n\mapsto\gg \\
        (\cdot)^\vee & :\gg\mapsto\RR^n.
    \end{aligned}
\end{equation}

The mapping between $\RR^n$ and $\Gg$ can be defined using the functions $\Exp(\cdot):\RR^n\mapsto\Gg$ and $\Log(\cdot):\Gg\mapsto\RR^n$ for any $\phi\in\RR^n$ and $\Xx\in\Gg$ as follows:
\begin{equation}
    \begin{aligned}
        \Exp(\phi) & =\expm(\phi^\wedge)=\Xx \\
        \Log(\Xx) & =\logm(\Xx)^\vee=\phi
    \end{aligned}
\end{equation}
where $\expm(\cdot)$ and $\logm(\cdot)$ are the matrix exponential and logarithm, respectively. 

The adjoint action, denoted as $\Ad_\Xx:\gg\mapsto\gg$ for any $\Xx\in\Gg$, is a Lie algebra isomorphism that allows change of frames. Given $\phi,\xi,\eta\in\RR^n$ and $\phi^\wedge,\xi^\wedge,\eta^\wedge\in\gg$, the adjoint action can be expressed in the function form as 
\begin{equation}
    \Ad_\Xx(\phi)=\Xx\phi^\wedge\Xx^{-1}
\end{equation}
or in the matrix form as 
\begin{equation}
    (\Ad_\Xx\phi)^\wedge=\Xx\phi^\wedge\Xx^{-1}.
\end{equation}

The adjoint map is the derivative of the adjoint action with respect to $\Xx$ at the identity element and is defined as 
\begin{equation}
    \ad_\xi\eta=[\xi^\wedge,\eta^\wedge]
\end{equation}
where  $[\xi^\wedge,\eta^\wedge]$ is the Lie bracket, defined as 
\begin{equation}
    [\xi^\wedge,\eta^\wedge]=\xi^\wedge\eta^\wedge-\eta^\wedge\xi^\wedge   .
\end{equation}

\section{Problem Definition}
\label{problem-definition}
We consider systems whose states reside in the tangent bundle of a matrix Lie group. This encompasses a diverse array of systems \cite{Kobilarov2011} whose states can be represented as pairs $\{\Xx, \xi^\wedge\} \in \Gg \times \gg$, where $\Xx$ represents a configuration and $\xi^\wedge$ represents the velocity (rate of change) of that configuration. The continuous-time equations of motion for such systems can be written as:
\begin{equation}
    \label{eq:lie-group-dyn}
    \begin{aligned}
        \dot{\Xx}_t&=\Xx_t\xi^\wedge_t\\
        \dot{\xi}_t&=f\big(\Xx_t,\xi_t,u_t\big)
    \end{aligned}
\end{equation}
where $u_t\in\RR^m$ is the generalized control input and $f(\cdot)$ is the function of velocity dynamics. For a given initial state $\{\Xx_0, \xi_0\}$, a goal state $\{\Xx_g, \xi_g\}$, and a time horizon $N$, we define the discrete-time constrained optimal control problem as
\begin{equation}
	\label{cddp_lie_problem}
	\begin{aligned}
		\underset{u_0, ..., u_{N-1}}{\text{min}} \quad & \ell^f(\Xx_N, \xi_N)+\sum_{k=0}^{N-1}\ell(\Xx_k, \xi_k, u_k)  \\
		\text{subject to} \quad & \Xx_{k+1} = F_{\Xx}(\Xx_k,\xi_k), \quad k=0,...,N-1\\ 
        & \xi_{k+1} = F_{\xi}(\Xx_k,\xi_k,u_k), \quad\;\;\; k=0,...,N-1 \\ 
		& u_{min} \leq u_k \leq u_{max}, \quad \; \forall k, \\ 
        & \bm{g}(\Xx_k,\xi_k, u_k)\leq \bm{0}, \quad \;\;\; \forall k, \\
		\text{given} \quad & \Xx_0, \xi_0,
	\end{aligned}
\end{equation}
where $\ell^f:\Gg\times\RR^n\mapsto\RR$ and $\ell:\Gg\times\RR^n\times\RR^m\mapsto\RR$ are the final cost and the running cost, respectively. $F_{\Xx}$ and $F_{\xi}$ are the discretized form of the configuration and velocity dynamics, which can be obtained by using a zero-order hold or first-order Euler integration method with a fixed time step $\Delta t$. Lastly, $\bm{g}$ is a vector of \textit{p} constraints in the form of differentiable nonlinear functions representing the state constraints.

\section{Proposed Method}
\label{sec:proposed-method}
The analytic solution to the general problem outlined in Section \ref{problem-definition} is often difficult. Additionally, finding the global minimum numerically can be time-consuming and likely infeasible, particularly for nonlinear systems with high state dimensions. Therefore, we propose a method that finds feasible solutions, even if they may not be globally optimal. To accomplish this, we utilize the \ac{ddp} framework \cite{Jacobson1970}, which iteratively solves sub-optimization problems in the backward pass and generates a new trajectory in the forward pass based on the found optimal policy, in order to approach a local optimum. 

To account for constraints imposed on the system, whose states lie in the tangent bundle of a matrix Lie group, we optimize an augmented Lagrangian function that combines trajectory cost with constraint penalties. The approach involves lifting the problem to the Lie algebra in the backward pass by computing the gradient of the cost function within the corresponding Lie algebra, and retracting back to the manifold in the forward pass by integrating the dynamics using the optimal policy obtained in the backward pass.

\subsection{Dynamics on Tangent Space}
The central concept of \ac{ddp} is that, at each iteration, all nonlinear constraints and objectives are approximated using first or second-order Taylor series expansions. This allows the approximate functions, which now operate on deviations from the nominal trajectory, to be solved using discrete \ac{lqr} techniques. In order to define the cost and constraint functions in Lie algebra, we need to determine the error dynamics for the configuration. To obtain the perturbed state dynamics, we followed the approach proposed by Teng~\etal \cite{Teng2022a}. For completeness, we outline the necessary steps here. Interested readers may refer to \cite{Teng2022a,Teng2022b} for more information.

Consider a perturbed state $\{\Xx_p, \xi^\wedge_p\}$ that is in the vicinity of a nominal state $\{\Xx, \xi^\wedge\}$. Then, the configuration error $\Psi$ can be defined as
\begin{equation}
    \label{configuration_error}
    \Psi=\Xx\inv\Xx_p\in\Gg.
\end{equation}
Differentiating both sides of (\ref{configuration_error}) yields the configuration error dynamics as
\begin{equation}
    \label{nonlinear_conf_err_dyn}
	\begin{aligned}
	\dot{\Psi} &=  \Xx\inv\frac{d}{dt}\Big(\Xx_p\Big)+\frac{d}{dt}\Big(\Xx\inv\Big)\Xx_p \\
	& = \Xx\inv\dot{\Xx_p}-\Xx\inv\dot{\Xx}\Xx\inv\Xx_p \\
	  & = \Xx\inv\Xx_p\xi_p^\wedge-\Xx\inv\Xx\xi^\wedge\Xx\inv\Xx_p \\
	  & = \Psi\xi_p^\wedge-\xi^\wedge\Psi.
	\end{aligned}
\end{equation}
Here, we can define a vector $\psi$ in $\RR^n$ such that the matrix exponential of $\psi^\wedge$ corresponds to $\Psi$, denoted as $\Psi = \expm(\psi^\wedge)$. Using the first-order approximation of the matrix exponential, which states that $\expm(\psi^\wedge) \approx I_n + \psi^\wedge$, the dynamics of the configuration error in (\ref{nonlinear_conf_err_dyn}) can be linearized as follows:
\begin{equation}
    \label{perturbed_conf_dyn}
    \begin{aligned}
    \dot{\Psi} & = \Psi\xi_p^\wedge-\xi^\wedge\Psi \\
    & \approx (I_n+\psi^\wedge)\xi_p^\wedge-\xi^\wedge(I_n+\psi^\wedge) \\
    & = \xi_p^\wedge+\psi^\wedge\xi_p^\wedge-\xi^\wedge-\xi^\wedge\psi^\wedge \\
    & = \xi_p^\wedge-\xi^\wedge+\psi^\wedge\xi_p^\wedge-\xi^\wedge\psi^\wedge \\
    & = \xi_p^\wedge-\xi^\wedge + \psi^\wedge(\xi^\wedge-\xi^\wedge+\xi_p^\wedge) - \xi^\wedge\psi^\wedge \\
    & = \xi_p^\wedge-\xi^\wedge + \psi^\wedge\xi^\wedge-\xi^\wedge\psi^\wedge + \psi^\wedge(\xi_p^\wedge-\xi^\wedge)\\
    & \approx \xi_p^\wedge-\xi^\wedge + \psi^\wedge\xi^\wedge-\xi^\wedge\psi^\wedge\\
    & = \xi_p^\wedge-\xi^\wedge + \ad_\psi\xi\\
    \dot{\psi} & = \xi_p^\wedge-\xi^\wedge-\ad_{\xi}\psi
    \end{aligned}
\end{equation}

Note that the second order term of $\psi^\wedge(\xi_p^\wedge-\xi^\wedge)$ is also discarded to obtain the linear dynamics of the configuration error \cite{Teng2022a,Teng2022b}. $\psi$ in (\ref{perturbed_conf_dyn}) is the perturbed configuration represented in Lie algebra. The perturbed velocity and control input are also defined as
\begin{equation}
    \delta \xi = \xi_p - \xi, \quad \text{and} \quad \delta u = u_p-u
\end{equation}

The perturbed velocity dynamics can then be approximated up to second order as:
\begin{equation}
    \label{perturbed_vel_dyn}
    \begin{aligned}
        \delta \dot{\xi}_t & = \bm{\Gamma}_t\delta \xi_t + \bm{\Lambda}_t\delta u_t \\ 
        &+ \frac{1}{2} \left( \bm{H}_{\xi_t} (\delta \xi_t \otimes \delta \xi_t) + \bm{H}_{u_t} (\delta u_t \otimes \delta u_t) \right) \\
        &+ \bm{H}_{\xi_t u_t} (\delta \xi_t \otimes \delta u_t) 
    \end{aligned}
\end{equation}
where $\bm{\Gamma}_t$ and $\bm{\Lambda}_t$ are the Jacobians, $\bm{H}_{\xi_t}$,$\bm{H}_{u_t}$, $\bm{H}_{\xi_t,u_t}$ are the Hessians of velocity dynamics ($F_{\xi}$) evaluated at the nominal trajectory. The tensor product $\otimes$ denotes the outer product of the perturbation vectors.

Lastly, we define the perturbed states (error-state as defined in \cite{Teng2022a}) as concatenation 
\begin{equation}
    \bm{x} = \begin{bmatrix}
    \psi\\
    \delta\xi
    \end{bmatrix}, \quad \bar{\bm{u}} = \delta u.
\end{equation}
Note that the perturbed state dynamics $\dot{\bm{x}} = h(\bm{x},\bm{u})$, are linear in terms of configuration states but nonlinear in terms of velocity dynamics. Ignoring the Hessian matrices, one can obtain linear perturbed state dynamics as
\begin{equation}
    \label{perturbed_lin_state_dyn}
        \dot{\bm{x}} =
    \underbrace{
    \begin{bmatrix}
    -\ad_{\xi} & I_n\\
    \bm{0}_{n\times n} & \bm{\Gamma}_t
    \end{bmatrix}
    }_{\triangleq\bm{A}_t}
    \bm{x} +
    \underbrace{
    \begin{bmatrix}
    \bm{0}_{n\times m}\\
    \bm{\Lambda}_t
    \end{bmatrix}
    }_{\triangleq\bm{B}_t}
    \bar{\bm{u}}
\end{equation}
Employing linear dynamics in \ac{ddp} framework refers to iterative LQR. 

The discrete-time counterparts of the matrices $\bm{A}_t$ and $\bm{B}_t$ can be simply obtained by applying a zero-order hold or a first-order Euler integration with a fixed time step $\Delta t$.

\subsection{Constraint Handling}
\label{subsec:const-handling}
In order to handle the constraints in \ac{ddp} framework, we approximate the constraints up to second-order as:
\begin{equation}
    \label{pert-state-constraints}
    \begin{aligned}
        \bar{\bm{c}}(\bm{x}+\delta\bm{x}, \bm{u}+\delta\bm{u}) & \approx \bar{\bm{c}}(\bm{x}, \bm{u}) \\
        & + \bar{\bm{c}}_{\bm{x}}\delta\bm{x} + \bar{\bm{c}}_{\bm{u}}\delta\bm{u}\\
         &+ \frac{1}{2}(\delta\bm{x}\Tra \bar{\bm{c}}_{\bm{xx}}\delta\bm{x}+\delta\bm{u}\Tra \bar{\bm{c}}_{\bm{uu}}\delta\bm{u}) \\
        &+ \delta\bm{x}\Tra \bar{\bm{c}}_{\bm{xu}}\delta\bm{u}.
    \end{aligned}
\end{equation}
where $\bar{\bm{c}}_{\bm{x}}$, $\bar{\bm{c}}_{\bm{u}}$ are the first, $\bar{\bm{c}}_{\bm{xx}}$,$\bar{\bm{c}}_{\bm{xu}}$, $\bar{\bm{c}}_{\bm{uu}}$ are the second derivatives of $\bar{\bm{c}}$ evaluated at nominal state trajectory. 

The difference between two group elements in the configuration state produces a geodesic in the group. To manage this, we propose mapping the distance geodesic to the tangent space of the configuration at the current time step and addressing the constraint in that vector space.

For instance, configuration avoidance constraints can typically be formulated as inequality constraints using an n-spherical function, with the center of the n-sphere located at the configuration to be avoided ($\Xx_c$) and the radius ($r_c$) defining the restricted region. This allows us to specify a region of configurations that should be avoided. The distance between the nominal and restricted configurations, $\Psi^c = \Xx\inv\Xx_c$, can be approximated in the tangent space of the nominal trajectory as:
\begin{equation}
\label{logm2ndapprox}
    \psi^c = \logm(\Psi^c) \approx ((\Psi^c-I)-0.5(\Psi^c-I)^2)^\vee 
\end{equation}
where $I$ is the identity matrix with the same size of $\Psi^c$. Then, the configuration avoidance constraint can be written as 
\begin{equation}
    \label{eq:Xx-constraint}
    \bar{\bm{c}}^{\Xx}(\psi^c) = (r_c^2-\|\psi^c\|_2)\leq 0
\end{equation}

In this approach, we consider the same restricted region for each axis in the n-dimensional sphere. However, it is also possible to specify different radius values for each axis, resulting in an n-dimensional ellipsoid as the restricted region. Our method can accommodate these types of configuration constraints as well.

\subsection{Augmented Lagrangian Formulation}

An effective method for solving constrained optimization problems is to incorporate the constraints in the objective function and iteratively increase the penalty for violating or approaching them. This technique, known as the penalty method, guarantees convergence to the optimal solution as the penalty coefficient approaches infinity. However, this may not be practical to implement in numerical optimization routines due to the limitations of finite precision arithmetic. Augmented Lagrangian methods \cite{Howell2019} offer an alternative solution by maintaining estimates of the Lagrange multipliers associated with the constraints, allowing for convergence to the optimal solution without requiring the penalty terms to increase infinitely. Here we obtain the augmented Lagrangian as 
\begin{equation}
    \label{eq:augmented-lagrangian}
    \begin{aligned}
        &\Ll_A  = \Ll_N(\bm{x}_N) + \sum_{k=0}^{N-1}\Ll_k(\bm{x}_k,\bm{u}_k) \\
        &\Ll_N(\bm{x}_N)  = \bar{\ell}^f(\bm{x}_N) + (\lambda_N+\frac{\mu}{2}\bar{\bm{g}}^f(\bm{x}_N)\odot\beta_N)^{\top}\bar{\bm{g}}^f(\bm{x}_N) \\
        &\Ll_k(\bm{x}_k,\bm{u}_k)  = \bar{\ell}(\bm{x}_k, \bm{u}_k) + (\lambda_k+\frac{\mu}{2}\bar{\bm{g}}(\bm{x}_k, \bm{u}_k)\odot\beta_k)^{\top}\bar{\bm{g}}(\bm{x}_k, \bm{u}_k)
    \end{aligned}
\end{equation}
where $\bar{\ell}^f:\RR^{2n}\mapsto\RR$ and $\bar{\ell}:\RR^{2n}\times\RR^m\mapsto\RR$ are the final cost and the running cost functions for perturbed system dynamics, respectively. A typical design of such functions is quadratic, 
\begin{equation}
    \begin{aligned}
        \bar{\ell}^f(\bm{x}) & = \frac{1}{2}\|\delta\bm{x}\|_{S_V} \\
        \bar{\ell}(\bm{x},\bm{u}) & = \frac{1}{2}\|\delta\bm{x}\|_{S_Q} + \frac{1}{2}\|\delta\bm{u}\|_{S_U}         
    \end{aligned},
\end{equation}
where $S_V\in\RR^{2n\times2n}$, $S_Q\in\RR^{2n\times2n}$ and $S_U\in\RR^{m\times m}$ are cost matrices that are specified by the user and remain constant.

In \eqref{eq:augmented-lagrangian}, $\bar{\bm{g}}^f$ and $\bar{\bm{g}}$ are vectors of \textit{p} constraints for final state and running state-inputs of perturbed state dynamics as introduced in (\ref{pert-state-constraints}), i.e., constraints represented in local tangent spaces. $\odot$ is element-wise multiplication operator, $\lambda_k\in\RR^p$ is the vector of Lagrange multipliers (dual variables), $\mu\in\RR$ is the penalty parameter and $\beta_k\in\RR^p$ is the binary control vector to indicate active constraints, defined as
\begin{equation}
    \beta_k[i]=
    \begin{cases}
    0 & \text{if } \bar{\bm{g}}_k[i]<0 \annd \lambda_k[i]\sim0,\\
    1 & \text{otherwise} 
    \end{cases}
\end{equation}
By doing so, satisfied constraints do not cause additional cost by penalty amount if their corresponding dual variable is close to zero. In general, the penalty parameter $\mu$ can vary over time, but we kept it constant for each time step within the same iteration for simplicity.

Augmented Lagrangian methods include an outer loop where the combined cost function (\ref{eq:augmented-lagrangian}) is formed for fixed $\lambda$ and $\mu$ parameters, transforming such a problem (\ref{cddp_lie_problem}) into an unconstrained one to be solved with an inner solver, such as DDP/iLQR. At each step of the outer loop, an increasing penalty is applied as follows:
\begin{equation}
\label{eq:lambda-mu-updates}
\begin{aligned}
\lambda_k^+ & =\max(0,\lambda_k+\mu\bar{\bm{g}}_k(\bm{x}_k^*,\bm{u}_k^*)) \quad \forall k\in\{0,...,N-1\} \\
\mu^+ & = \gamma\mu, \quad \gamma>0
\end{aligned}
\end{equation}
where $\{(\bm{x}_0^*,\bm{u}_0^*),...,(\bm{x}_{N-1}^*,\bm{u}_{N-1}^*)\}$ is the optimal state and input trajectory obtained from the inner solver and $\gamma$ is a fixed penalty update rate. The outer loop typically continues if the number of iterations is less than a specified maximum and the constraints have not yet been satisfied within the desired tolerance threshold. Constraint satisfaction is typically tested via
\begin{equation}
\begin{aligned}
    \texttt{MCV} = & \; \max_{k = 0, \ldots, N-1} \Big(\max(0, \bar{\bm{g}}_k(\bm{x}_k^*,\bm{u}_k^*))\Big) \\
    \texttt{MCD} = & \; \max_{k = 0, \ldots, N-1} \Big(|\bar{\bm{g}}_k(\bm{x}_k^*,\bm{u}_k^*)\odot\lambda_k|\Big)
\end{aligned}
\end{equation}
where if either $\texttt{MCV}$ (maximum constraint violation) or $\texttt{MCD}$ (maximum complementary dual) is higher than a small threshold $\varepsilon_c$, further improvement in constraint satisfaction is needed, so the outer loop continues.
\begin{equation}
    \label{eq:ConsViolCheck}
    \text{\textit{ConsViolCheck}}=
    \begin{cases}
    1 & \text{if } \texttt{MCV}>\varepsilon_c \vee \texttt{MCD}>\varepsilon_c,\\
    0 & \text{otherwise} 
    \end{cases}
\end{equation}
The entire algorithm is described in Algorithm \ref{eq:CDDP-algorithm}.

\subsection{Differential Dynamic Programming}

Given augmented Lagrangian function (\ref{eq:augmented-lagrangian}) from outer loop, we define the cost-to-go and action-value functions $V$ and $Q$ as
\begin{equation}
\label{eq:Q_and_V}
\begin{aligned}
    V_k(\bm{x}_k) & =\underset{\bm{u}_k}{\min}\{\Ll_k(\bm{x}_k,\bm{u}_k)\}+V_{k+1}(\bm{A}_k\bm{x}_k+\bm{B}_k\bm{u}_k) \\
    &=\underset{\bm{u}_k}{\min}\;Q(\bm{x}_k,\bm{u}_k)).
\end{aligned}
\end{equation}
The matrices $\bm{A}_k$ and $\bm{B}_k$ represent the discretized versions of $\bm{A}_t$ and $\bm{B}_t$ in equation (\ref{perturbed_lin_state_dyn}). 
A second-order Taylor series expansion of the cost-to-go function can be written 
\begin{equation}
    \delta V_k(x) \approx \frac{1}{2}\delta\bm{x}_k^{\top}V_{xx,k}\delta\bm{x}_k + V_{x,k}^{\top}\delta\bm{x}_k
\end{equation}
where $V_{xx,k}$ and $V_{x,k}$ are the Hessian and gradient of the cost-to-go at time step $k$, respectively. The action-value function defined in \eqref{eq:Q_and_V} can be also approximated as a quadratic function as
\begin{equation}
    \label{eq:quadratic-Q}
    \begin{aligned}
        Q(\bm{x}+\delta\bm{x},\bm{u}+\delta\bm{u}) &\approx Q(\bm{x},\bm{u}) + Q_{\bm{x}}\Tra\delta\bm{x} +  Q_{\bm{u}}\Tra\delta\bm{u} \\ 
        &+ \frac{1}{2}(\delta\bm{x}\Tra Q_{\bm{xx}}\delta\bm{x}+\delta\bm{u}\Tra Q_{\bm{uu}}\delta\bm{u}) \\
        &+ \delta\bm{u}\Tra Q_{\bm{ux}}\delta\bm{x}.
    \end{aligned}
\end{equation}

To compute the derivative matrices in (\ref{eq:quadratic-Q}):
\begin{equation}
    \label{eq:Q-derivatives}
    \begin{aligned}
        Q_{\bm{x}} & = \Ll_{\bm{x}}+\bm{A}\Tra V_{x}\Next \\
        Q_{\bm{u}} & = \Ll_{\bm{u}}+\bm{B}\Tra V_{x}\Next \\
        Q_{\bm{xx}} & = \Ll_{\bm{xx}}+\bm{A}\Tra V_{xx}\Next\bm{A}+ V_{x}\Next h_{\bm{xx}} \\
        Q_{\bm{uu}} & = \Ll_{\bm{uu}}+\bm{B}\Tra V_{xx}\Next\bm{B}+V_{x}\Next h_{\bm{uu}} \\
        Q_{\bm{ux}} & = \Ll_{\bm{ux}}+\bm{B}\Tra V_{xx}\Next\bm{A}+V_{x}\Next h_{\bm{ux}} \\
    \end{aligned}
\end{equation}

To simplify the notation, we have omitted the time indices on all variables. All variables in this expression are evaluated at time step $k$, except for those marked with $\Next$, which are evaluated at time step $k+1$.

Calculating the full second-order expansion of the state dynamics ($h_{\bm{xx}}$, $h_{\bm{uu}}$, $h_{\bm{ux}}$), can be computationally expensive, particularly for systems with complex dynamics and high-dimensional states. In such cases, employing a first-order approximation, as done in iterative LQR, can be an alternative. However, while this approach reduces computational cost by providing a Gauss-Newton approximation of the true Hessian, it also diminishes local fidelity and increases the number of iterations required for convergence.

Minimizing (\ref{eq:quadratic-Q}) with respect to $\delta\bm{u}$ results in an affine controller 
\begin{equation}
    \label{eq:opt-control-policy}
    \delta\bm{u}^*=-Q_{\bm{uu}}^{-1}(Q_{\bm{ux}}\delta\bm{x}+Q_{\bm{u}})\triangleq \bm{K}\delta\bm{x}+\bm{d}.
\end{equation}

In cases of rank deficient $Q_{\bm{uu}}$, it is typically regularized as $Q_{\bm{uu}}+\rho I$, where $\rho$ is a small positive constant and $I$ is the identity matrix. Substituting $\delta\bm{u}^*$ into (\ref{eq:quadratic-Q}) yields the derivatives of the cost-to-go at time step $k$ in terms of the derivatives of the action value function as
\begin{equation}
\label{eq:V-derivatives}
    \begin{aligned}
        V_{\bm{x}} & = Q_{\bm{x}}+\bm{K} Q_{\bm{u}} + \bm{K}\Tra Q_{\bm{uu}}\bm{d}+Q_{\bm{ux}}\Tra\bm{d},\\
        V_{\bm{xx}} & = Q_{\bm{xx}}+\bm{K}\Tra Q_{\bm{uu}}\bm{K}+\bm{K}\Tra Q_{\bm{ux}}+Q_{\bm{ux}}\Tra\bm{K}.
    \end{aligned}
\end{equation}

At the final time step, $V_{\bm{x}}$ and $V_{\bm{xx}}$ can be easily computed as the first and second derivatives of the final Lagrangian function ($\Ll_N$). This way, the derivatives of the action-value function \eqref{eq:Q-derivatives} and in turn the local optimal control policy \eqref{eq:opt-control-policy} at each step can be calculated backwards starting from the final step. 

\begin{algorithm}[t!]
\begin{algorithmic}[1]
\State \textbf{Given:} $\Xx_0, \xi_0, \Xx_g, \xi_g, N$
\State \textbf{Init:} $\bm{U}=\{\bm{u}_0,...,\bm{u}_{N-1}\}$ 
\State \textbf{Calculate:} $\{\Xx_1, \xi_1, ... \Xx_N, \xi_N\}$ integrating \eqref{eq:lie-group-dyn}
\State \textbf{Calculate:} $\bm{X}=\{\bm{x}_0,...,\bm{x}_{N}\}$ error-states between $\Xx_i,\xi_i$ and $\Xx_g,\xi_g$
\State \textbf{Init:} $\lambda=0, \mu, iter_{max}^{al},iter_{max}^{ddp}, it_{al}=0,it_{ddp}=0$
\State $J \leftarrow \Ll_A(\bm{X},\bm{U};\lambda,\mu)$ (\ref{eq:augmented-lagrangian})
\While {\text{\textbf{not} \texttt{AL\_Converged} }}
    \State $\texttt{DDP\_Converged}$ = False
    \While {\text{\textbf{not} \texttt{DDP\_Converged} }}
    \State $\bm{K},\bm{d} \leftarrow$ Backward Pass (Algorithm \ref{DDP_BP})
    \State $\bm{X}^{\dagger},\bm{U}^{\dagger}, J^{\dagger} \leftarrow$ Forward Pass (Algorithm \ref{DDP_FP})
    \If{$(it_{ddp}<iter_{max}^{ddp}) \; \texttt{and} \\ \; (||\bm{U}-\bm{U}^{\dagger}||> 0) \; \texttt{and} \; (J^{\dagger}-J<0)$}
        \State $\bm{X}, \bm{U}, J \leftarrow \bm{X}^{\dagger}, \bm{U}^{\dagger}, J^{\dagger}$
        \State $it_{ddp} \leftarrow it_{ddp}+1$
    \Else
        \State $\texttt{DDP\_Converged}$ = True 
    \EndIf
    \EndWhile
    \If{$(it_{al}<iter_{max}^{al}) \; \texttt{and} \; $(ConsViolCheck$(\ref{eq:ConsViolCheck}))$}
        \State Update $\lambda, \mu$ (\ref{eq:lambda-mu-updates})
        \State $it_{al} \leftarrow it_{al}+1$
    \Else
        \State $\texttt{AL\_Converged}$ = True
        \State $\bm{X}^*, \bm{U}^* \leftarrow \bm{X}, \bm{U}$
        \State \textbf{return} $\bm{X}^*, \bm{U}^*$
    \EndIf
\EndWhile
\caption{Constrained DDP for Matrix Lie Groups\label{eq:CDDP-algorithm}}
\end{algorithmic}
\end{algorithm}

\begin{algorithm}[t!]
\begin{algorithmic}[1]
\State \textbf{Given:} $\{\bm{x}_0,...,\bm{x}_{N}\}, \{\bm{u}_0,...,\bm{u}_{N-1}\} $
\For{\textnormal{\textbf{parallel}} $i=0,...,n_{\rho}$}
    \State $\rho_i\in\Pp$
    \Function{Backward\_Pass}{}
        \State $V_{\bm{x}}(\bm{x}_N), V_{\bm{xx}}(\bm{x}_N) \leftarrow \frac{\mathstrut\partial\Ll_N}{\mathstrut\partial\bm{x}}(\bm{x}_N), \frac{\mathstrut\partial^2\Ll_N}{\mathstrut\partial\bm{x}^2}(\bm{x}_N)$
        \For{$k=N-1, ... ,0$}
            \State Calculate the derivatives of Lagrangian (\ref{eq:augmented-lagrangian}) $\Ll_{\bm{x}}, \Ll_{\bm{u}}, \Ll_{\bm{ux}}, \Ll_{\bm{xx}}, \Ll_{\bm{uu}}$
            \State Calculate the derivatives of perturbed state dynamics (\ref{perturbed_conf_dyn}), (\ref{perturbed_vel_dyn}) $\bm{A}, \bm{B}, h_{\bm{xx}}, h_{\bm{ux}}, h_{\bm{uu}}$
            \State Calculate the derivatives of $Q$ \eqref{eq:Q-derivatives}
            \State $Q_{\bm{uu}}\leftarrow Q_{\bm{uu}}+\rho_i I$
            \State Calculate $\bm{K}_k^i,\bm{d}_k^i$ (\ref{eq:regularized-K-d})
        \EndFor 
        \State \textbf{return} $\bm{K}^i,\bm{d}^i=\{\bm{K}_k^i,\bm{d}_k^i | k=0,...,N-1\}$
    \EndFunction
    \State \textbf{return} $\bm{K},\bm{d}=\{\bm{K}^i,\bm{d}^i | i=0,...,n_{\rho}\}$
\EndFor
\caption{Backward Pass \label{DDP_BP}}
\end{algorithmic}
\end{algorithm}

\begin{algorithm}[t!]
\begin{algorithmic}[1]
\State \textbf{Given:} $\Xx_0, \xi_0, \Xx_g, \xi_g, \{\bm{K}^i,\bm{d}^i | i=0,...,n_{\rho}\}$
\For{\textnormal{\textbf{parallel}} $i=0,...,n_{\rho}$}
    \Function{Forward\_Pass}{}
        \State $\bar{\Xx}_0, \bar{\xi}_0 \leftarrow \Xx_0, \xi_0$
        \For{$k=0, ... , N-1$}
            \State Calculate $\delta\bm{x}_k$ in local tangent space \eqref{eq:state-error-in-tangent}
            \State Calculate $\bar{\bm{u}}_k^i$ \eqref{eq:updated-input} with $\bm{K}^i,\bm{d}^i,\alpha=1$
            \State Calculate $\bar{\Xx}_{k+1}, \bar{\xi}_{k+1}$ on manifold \eqref{eq:integrate-on-manifold}
            \State Store error-state $\bm{x}_{k+1}^i$ between $\bar{\Xx}_{k+1}, \bar{\xi}_{k+1}$ and $\Xx_{g}, \xi_{g}$
        \EndFor 
        \State $\bm{X}^i,\bm{U}^i=\{\bm{x}_0^i,...,\bm{x}_{N}^i\}, \{\bm{u}_0^i,...,\bm{u}_{N-1}^i\}$
        \State Calculate trajectory cost $J^i$ using (\ref{eq:augmented-lagrangian})
        \State \textbf{return} $\bm{X}^i,\bm{U}^i, J^i$
    \EndFunction
    \State $i^* = \texttt{argmin} \{J^i | i=0,...,n_{\rho}\}$
    \State \textbf{return} $ \bm{X}^{i^*},\bm{U}^{i^*}, J^{i^*} $
\EndFor
\caption{Forward Pass \label{DDP_FP}}
\end{algorithmic}
\end{algorithm}

After determining the optimal control policy for each time step, we update the nominal trajectories by simulating the dynamics forward on the manifold itself starting from the initial state as:
\begin{equation}
\label{eq:state-error-in-tangent}
    \begin{aligned}
        \delta\bm{x}_k & = \begin{bmatrix}
        \logm(\Xx_k\inv\bar{\Xx}_k)\\
        \bar{\xi}_k-\xi_k
        \end{bmatrix}
    \end{aligned}
\end{equation}
\begin{equation}
\label{eq:updated-input}
    \begin{aligned}
        \delta\bm{u}_k & = \bm{K}_k\delta\bm{x}_k+\alpha\bm{d}_k \\
        \bar{\bm{u}}_k & = \bm{u}_k+\delta\bm{u}_k 
    \end{aligned}
\end{equation}
\begin{equation}
\label{eq:integrate-on-manifold}
    \begin{aligned}
        \bar{\xi}_{k+1} & = \bar{\xi}_k + f(\xi, \bar{\bm{u}}_k)\Delta t \\
        \bar{\Xx}_{k+1} & = \bar{\Xx}_k\expm(\bar{\xi}_{k+1}\Delta t)
    \end{aligned}
\end{equation}
where $\{\Xx_k,\xi_k,u_k\}$ and $\{\bar{\Xx}_k,\bar{\xi}_k,\bar{u}_k\}$ represent the nominal state-actions and the updated state-actions at time step $k$, respectively. In \eqref{eq:updated-input}, $\alpha$ is a scaling term for a simple linear search on the feedforward term. 
 
The convergence of DDP/iLQR methods is highly dependent on the search direction and step size \cite{Roulet2022}, which can be regularized by $\rho$ and $\alpha$, respectively. Although there is no proven optimal approach to determine the best search direction and step size for such constrained nonlinear optimization problems, there are promising practices such as gradually increasing regularization and employing Armijo line search strategies \cite{Howell2019,Xie2017}. 

In this work, we generate multiple controllers in parallel by varying $\rho$ during the backward pass (Algorithm \ref{DDP_BP}) as follows:
\begin{equation}
    \label{eq:regularized-K-d}
    \begin{aligned}
        \bm{K}^i = & -(Q_{\bm{uu}}+\rho_i I)\inv Q_{\bm{ux}} \\   
        \bm{d}^i = & -(Q_{\bm{uu}}+\rho_i I)\inv Q_{\bm{u}}
    \end{aligned}
\end{equation}
where we select $\rho_i$ from the set $\Pp$. With a small $\rho$, the regularization term has minimal impact, and the control updates are similar to the unregularized \textit{aggressive} updates, heavily relying on the information from $Q_{\bm{uu}}$. In contrast, with a large $\rho$, the control updates become more \textit{conservative} as the regularization term dominates, ensuring better numerical stability but slower convergence. This behavior is analogous to the Levenberg-Marquardt method, which transitions from the Gauss-Newton method to gradient descent as the regularization increases.

Once we obtain multiple controllers in the backward pass ranging from aggressive to conservative, we rollout the trajectories in closed loop with each controller and $\alpha$$=$1 in the forward pass and calculate the cost (Algorithm \ref{DDP_FP}). The trajectory with the minimum cost is taken as the nominal trajectory for the next iteration of DDP, if it also improves upon the previous iteration.

\subsection{Implementation Practices}
The proposed method is implemented in Python using the \texttt{jax} library \cite{jax} due to its capabilities for easy parallelization and automatic differentiation. The Jacobian and Hessian functions of the perturbed state dynamics, along with the first and second derivatives of the Lagrangian function, are obtained via \texttt{jax}'s automatic differentiation. The initial penalty parameter, $\mu$, determines whether the algorithm starts unconstrained (with a small $\mu$ value) or constrained and feasible (with a high $\mu$ value). Typically, we achieved a balance by setting $\mu$$=$1 and iteratively multiplying it by $\gamma$$=$10. The constraint satisfaction tolerance threshold ($\varepsilon_c$) is set to 1e$-$3. For varying regularization in the backward pass, we use $n_{\rho}$$=$$7$ and $\Pp$$=$$\{\rho_i=10^{i-4}|i=0,...,n_{\rho}\}$, which generates varying search directions and control updates from \textit{aggressive} ($\rho_0$$=$1e$-$4) to \textit{conservative} ($\rho_7$$=$1000).

\section{Experiments}
\label{sec:Experiments}

The proposed method is employed here to optimize the trajectory of a system, whose states lie in the tangent bundle of a matrix Lie group. Experiments\footnote{The videos of the experiments can be found on the project website:\\ \texttt{https://sites.google.com/view/cddp-lie}} were conducted to evaluate the method based on:

\begin{itemize}
\item Handling constraints in a simple mechanical system characterized by rigid body dynamics in SE(3),
\item Mitigating external disturbances through its application as a Lie-algebraic feedback control policy,
\item Computational efficiency compared to existing optimization solvers,
\item Trajectory optimization performance in a realistic quadrotor scenario.
\end{itemize}

\subsection{Tests with SE(3) Dynamics}
We consider a 3D rigid body in SE(3) where the states of the system can be represented by a rotation matrix
\begin{equation}
    R \in \text{SO(3)} \equiv
 \{R\in\RR^{3\times3}|R^{\top}R=I_3,\det(R)=1\}
\end{equation}
and position $p\in\RR^3$. The homogeneous representation of a group element in SE(3) is
\begin{equation}
    \Xx=\begin{bmatrix}
    R & p\\
    0 & 1
    \end{bmatrix} \in \text{SE(3)}
\end{equation}
The velocity vector $\xi$ in SE(3) is known as a ``twist'' and consists of both angular ($\omega$) and linear ($v$) velocities in body frame as 
\begin{equation}
    \xi=\begin{bmatrix}
    \omega \\
    v
    \end{bmatrix}\in \RR^6,
    \quad \xi^\wedge=
    \begin{bmatrix}
    \omega^\wedge & v \\
    0 &0
    \end{bmatrix} \in \se(3)
\end{equation}
The forced Euler-Poincaré equations \cite{Bloch1996} define the twist dynamics as
\begin{equation}
    J_b\dot{\xi}=\ad^*_{\xi}J_b\xi+u
\end{equation}
In this expression, $J_b$ represents the generalized inertia matrix in the body-fixed principal axes, while $u \in \gg^*$ represents the generalized control inputs including torques and forces ($u_{\tau}$ and $u_f$) applied to these axes. $\gg^*$ denotes the cotangent space and the coadjoint map is represented by $\ad_{\xi}^*$. The matrix representations of the adjoint action in the Lie algebra ($\ad_{\xi}$) and the coadjoint map ($\ad^*_{\xi}$) are as follows:
\begin{equation}
    \ad_{\xi}=\begin{bmatrix}
    \omega^\wedge & 0 \\
    v^\wedge & \omega^\wedge
    \end{bmatrix}, \quad 
    \ad^*_{\xi}=\ad^{\top}_\xi=-\begin{bmatrix}
    \omega^\wedge & v^\wedge \\
    0 & \omega^\wedge
    \end{bmatrix}.
\end{equation}
Lastly, the continuous equations of motion described in \eqref{eq:lie-group-dyn} can be written for SE(3) group as
\begin{equation}
    \begin{aligned}
        \dot{\Xx} & = \Xx\xi^\wedge, \\
        \dot{\xi} & = J_b\inv \Big( \ad_{\xi}^*J_b\xi+u \Big)
    \end{aligned}
\end{equation}

\subsubsection{Constraint Handling} \label{subsec-const-handling}
We now test the proposed algorithm for planning a safe path for a rigid body in SE(3). The task involves rotating the body from the identity to the configuration $R_z(170^\circ)$ and translating it from the initial position $(2,2,2)$ to $(6,2,2)$ within 6 seconds, using a fixed time step of $\Delta t$$=$0.1. Rotations around the body-fixed frame's $x$, $y$, and $z$-axes are denoted by $R_x(\cdot)$, $R_y(\cdot)$, and $R_z(\cdot)$, respectively. 

During the motion, the configuration $R_z(90^\circ)$ is considered unsafe and must be avoided. Additionally, there is a spherical obstacle at $(4,2,2)$ with a radius of 1 that must also be avoided. All inputs were restricted between -7.5 and 7.5 and angular velocities were bounded at 0.5 when the position $x$ was less than 3; otherwise, the upper bound becomes 2.0. These constraints were formulated for all discrete steps, $k$, as follows: 
\begin{equation}
    \label{se3-constraints}
    \begin{aligned}
    \text{Obstacle Avoid: } & \; ||p_k-(4,2,2)||_2\geq 1.0 \\
    \text{Conf. Avoid: } & \;||\logm (R_{z}(90)\inv R_k)||_2 \geq 0.4 \\
    \text{Input Limits: } & \; -7.5 \leq u_k \leq 7.5 \\
    \text{Velocity Bound: } & \; \omega_{k} \leq \bar{\omega}_k \; \\ 
     & \; \bar{\omega}_k
\begin{cases}
    0.5, & \text{if } x_k \leq 3\\
    2.0,              & \text{otherwise}
\end{cases} \\
    \end{aligned}
\end{equation}
where $\logm$ operation in configuration avoidance constraint was implemented with its second order approximation as described in (\ref{logm2ndapprox}).  

\begin{figure}[t!]
	\centering
	\def\svgwidth{0.95\linewidth}
	{
 \fontsize{7}{7}
		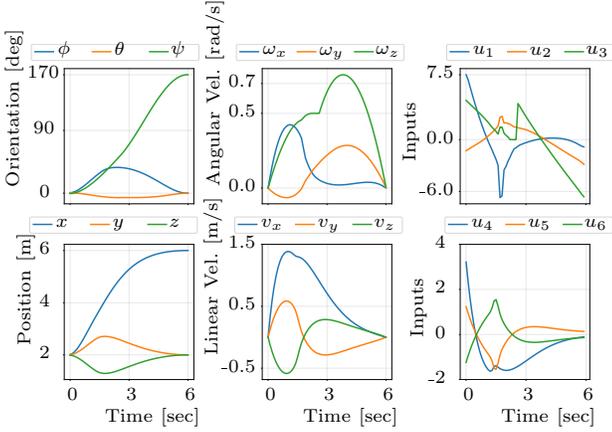}
	\caption{The proposed constrained DDP method effectively optimizes the trajectory for an SE(3) object, taking into account constraints on position, orientation, velocities and inputs.}
	\label{fig:se3-example}
\end{figure}

The configuration trajectories were converted to Euler angles ($\phi, \theta, \psi$ in degrees) using the rotation order $R_x(\cdot)R_y(\cdot)R_z(\cdot)$. In resultant trajectories, rotations around the $x$ and $y$-axes (changes in $\phi$ and $\theta$ angles) were observed to avoid the unsafe configuration of $R_x(90^\circ)$ (Fig.~\ref{fig:se3-example}). When accounting for the configuration constraint, the robot changed its motion along different axes to bypass the unsafe orientation. The presence of the spherical obstacle required a more circuitous route, deviating along each axis to achieve a collision-free trajectory (Fig.~\ref{fig:se3-example}). Furthermore, the angular velocity $\omega_z$ was saturated at 0.5 between 2-2.5 seconds, satisfying the position-dependent velocity constraint.

\subsubsection{Disturbance Rejection}
To evaluate the effectiveness of the proposed \ac{ddp} method in handling external disturbances, we extend the analysis done in \cite{Boutselis2020} to SE(3) and employ the proposed \ac{ddp} method as a Lie-algebraic feedback control:
\begin{equation}
    \label{eq:fb-policy}
    u_{k}^{fb} := u_k^*+\bm{K}_k\logm((\Xx_k^{*})\inv\Xx^{\epsilon}_k) 
\end{equation}
where $u_k^*$ and $\bm{K}_k$ represent the (sub)optimal control sequence and time-varying feedback gains, respectively, which are obtained through \ac{ddp}. The feed-forward term $\bm{d}_k$ in \eqref{eq:opt-control-policy} is not explicitly shown in the feedback policy \eqref{eq:fb-policy} because it is already included in $u_k^*$. The variables $\Xx_k^{*}$ and $\Xx^{\epsilon}_k$ represent the (sub)optimal states obtained through \ac{ddp} and the perturbed states due to disturbance, respectively.

\begin{figure}[t!]
	\centering
	\def\svgwidth{\linewidth}
	{
 \fontsize{9}{9}
		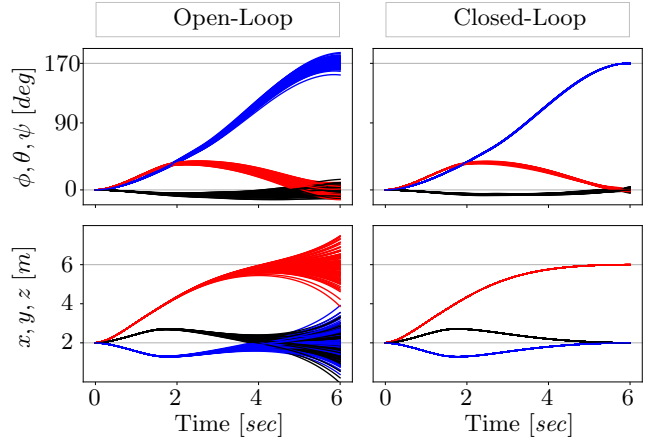}
	\caption{The implementation of Lie-algebraic feedback control in the proposed \ac{ddp} method notably reduces state variance, especially in the vicinity of goal points, when subjected to random disturbances.\label{fig:noise-test}}
\end{figure}

We test the policy (\ref{eq:fb-policy}) on the following stochastic version of the SE(3) dynamics:
\begin{equation}
    \begin{aligned}
        \Xx_{k+1}^{\epsilon} & = \Xx_k^{\epsilon}\expm_{SE(3)}(\xi_k^{\epsilon}\Delta t) \\
        \xi_{k+1}^{\epsilon} & = \xi_{k}^{\epsilon} + f(\xi_{k}^{\epsilon}, u_k)\Delta t + \sigma_{w}w
    \end{aligned}
\end{equation}
where $w$ is assumed to be spatially uncorrelated independent and identically distributed noise, drawn from a zero mean Gaussian distribution, $w\sim \mathcal{N}(0_{6\times1},\,I_6)$, $\sigma_{w}=0.001$. The performance of the proposed \ac{ddp} method under stochastic conditions was evaluated by allowing the optimizer to converge on the deterministic system and then testing its performance on the stochastic system. Fig.~\ref{fig:noise-test} shows the results of this comparison, using 1000 sampled trajectories under noisy dynamics to compare the open-loop policy $u_k^*$ with the feedback policy $u_k^{fb}$ \eqref{eq:fb-policy}. The results demonstrate that the use of the obtained feedback gains significantly reduces state variance, particularly in the vicinity of the goal points, as expected.

\subsubsection{Computational Efficiency}
A widely used approach to solving discrete optimal control problems involves utilizing readily available optimization solvers, which can be seamlessly applied with minimal modifications. Generating feasible trajectories can be achieved by treating the system dynamics as equality constraints and formulating both state and input constraints as inequality constraints.

In order to establish a benchmark comparison, we employed a Python optimizer that uses the Sequential Least SQuares Programming (SLSQP) Algorithm \cite{Kraft1988}, from the SciPy Optimization module and Interior Point Optimizer (IPOPT) \cite{Byrd1999}. The task described in Section \ref{subsec-const-handling} was tested by those methods. To accelerate convergence, we supplied derivative information for the cost function as well as equality and inequality constraints. 

\begin{figure}[t!]
	\centering
	\def\svgwidth{0.95\linewidth}
	{
 \fontsize{7}{7}
		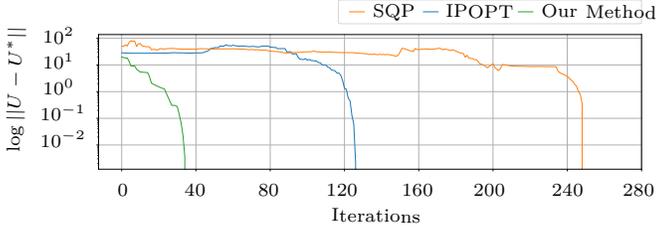}
	\caption{Input convergence comparison between standard constrained IPOPT, SQP and proposed constrained \ac{ddp} methods, showcasing the significantly faster convergence achieved by our method.}
	\label{fig:comparison}
\end{figure}

Figure \ref{fig:comparison} illustrates a comparison of convergence between the standard constrained IPOPT, SQP, and the proposed constrained DDP methods. DDP achieves convergence in less than 40 iterations, whereas IPOPT and SQP require 125 and 250 iterations, respectively. The primary reason for this difference is that direct optimization methods increase the number of decision variables as the horizon expands, leading to a search in a space with numerous equality and inequality constraints. 

\subsection{Trajectory Optimization of a Quadrotor}

We now consider the dynamics of a quadrotor as a 3D rigid body where the states of the system are represented by a rotation matrix $R\in SO(3)$, a position vector $p\in\RR^3$, an angular velocity vector $\omega\in\RR^3$ and a linear velocity vector $v\in\RR^3$ \cite{Alcan2017}. Unlike SE(3) dynamics, a quadrotor is an under-actuated robotic system, meaning it lacks generalized control inputs to directly control all six degrees of freedom. Instead, it is actuated by four motors, whose speeds ($P_i$, $i=0,1,2,3$) are assumed to be controlled nearly instantaneously \cite{Panerati2021}, inducing torques and forces as follows:
\begin{equation}
    \label{eq:ind-tor-for}
    \begin{aligned}
    \tau_{in} = & \begin{bmatrix}
    (P_1^2-P_3^2) k_F L \\
    (P_2^2-P_0^2) k_F L \\
    \sum_{i=0}^3(-1)^{i+1}k_T P_i^2
    \end{bmatrix}        
     \\ 
    F_{in} = &  \; R.[0,0,k_F\sum_{i=0}^3P_i^2] \\ 
    \end{aligned}
\end{equation}
where $L$ is the arm length and $k_F, k_T$ are constants. Here, we consider the quadrotor in a $+$ configuration, where its $x$ and $y$ axes are aligned with the motor locations. The induced torques and forces in (\ref{eq:ind-tor-for}) consequently generate linear acceleration in the global frame and changes in angular velocities in the local frame as:
\begin{equation}
    \begin{aligned}
    \dot{v} = & \; m\inv\Big( F_{in}-[0,0,mg] \Big) \\
    \dot{\omega} = & \; J\inv\Big(\tau_{in}-\omega\times(J\omega)\Big) \\
    \end{aligned}
\end{equation}
where $m,g$ and $J$ are mass, gravitational acceleration and inertia, respectively. Lastly, the position and orientation dynamics of the quadrotor in SE(3) are
\begin{equation}
    \dot{p} = v, \quad \dot{R} = R\omega^\wedge.
\end{equation}

\begin{figure}[b!]
\vspace{0.2cm}
	\centering
	\def\svgwidth{\linewidth}
	{
 \fontsize{7}{7}
		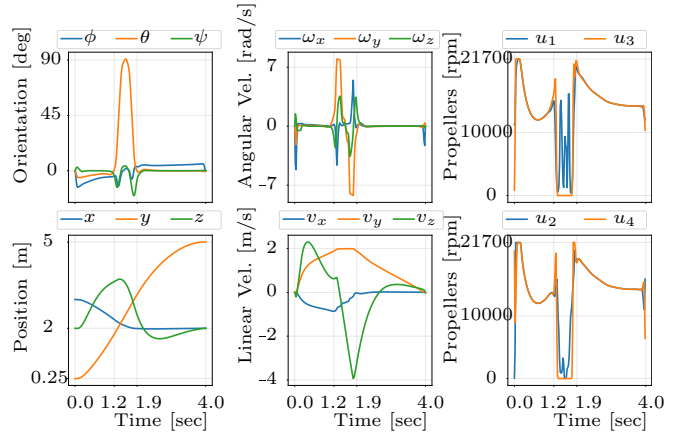}
	\caption{The proposed method successfully optimizes a trajectory for nonlinear quadrotor dynamics that requires a pure 90-degree rotation around the y-axis while avoiding constraints and without suffering from gimbal lock.}
	\label{fig:quadrotor-ex}
\end{figure}

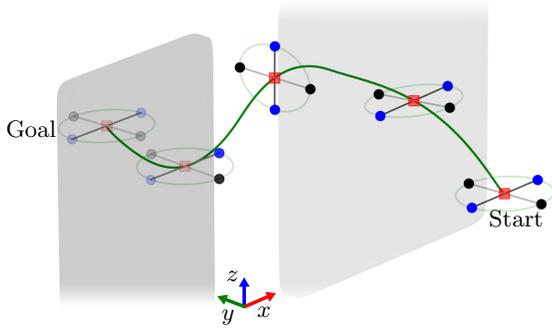
\begin{figure}[t!]
\vspace{0.2cm}
	\centering
	\def\svgwidth{0.9\linewidth}
	{
 \fontsize{9}{9}
\begingroup%
  \makeatletter%
  \providecommand\color[2][]{%
    \errmessage{(Inkscape) Color is used for the text in Inkscape, but the package 'color.sty' is not loaded}%
    \renewcommand\color[2][]{}%
  }%
  \providecommand\transparent[1]{%
    \errmessage{(Inkscape) Transparency is used (non-zero) for the text in Inkscape, but the package 'transparent.sty' is not loaded}%
    \renewcommand\transparent[1]{}%
  }%
  \providecommand\rotatebox[2]{#2}%
  \newcommand*\fsize{\dimexpr\f@size pt\relax}%
  \newcommand*\lineheight[1]{\fontsize{\fsize}{#1\fsize}\selectfont}%
  \ifx\svgwidth\undefined%
    \setlength{\unitlength}{533.99484709bp}%
    \ifx\svgscale\undefined%
      \relax%
    \else%
      \setlength{\unitlength}{\unitlength * \real{\svgscale}}%
    \fi%
  \else%
    \setlength{\unitlength}{\svgwidth}%
  \fi%
  \global\let\svgwidth\undefined%
  \global\let\svgscale\undefined%
  \makeatother%
  \begin{picture}(1,0.59059993)%
    \lineheight{1}%
    \setlength\tabcolsep{0pt}%
    \put(0,0){\includegraphics[width=\unitlength,page=1]{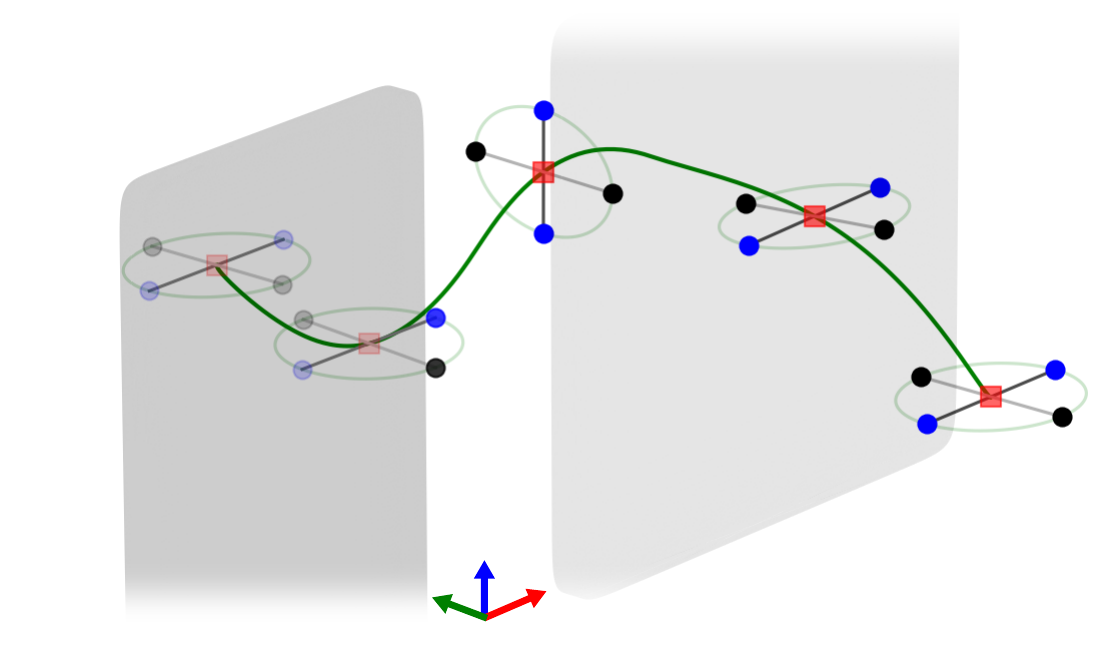}}%
    \put(0.457812,0.01794276){\makebox(0,0)[lt]{\lineheight{1.25}\smash{\begin{tabular}[t]{l}$x$\end{tabular}}}}%
    \put(0.39582162,0.01269278){\makebox(0,0)[lt]{\lineheight{1.25}\smash{\begin{tabular}[t]{l}$y$\end{tabular}}}}%
    \put(0.4041162,0.07961416){\makebox(0,0)[lt]{\lineheight{1.25}\smash{\begin{tabular}[t]{l}$z$\end{tabular}}}}%
    \put(0.86427908,0.17633511){\makebox(0,0)[lt]{\lineheight{1.25}\smash{\begin{tabular}[t]{l}$\text{Start}$\end{tabular}}}}%
    \put(0.01811498,0.33315771){\makebox(0,0)[lt]{\lineheight{1.25}\smash{\begin{tabular}[t]{l}$\text{Goal}$\end{tabular}}}}%
  \end{picture}%
\endgroup%
}
	\caption{The proposed method demonstrates an optimized trajectory for a quadrotor, showcasing a smooth 90-degree rotation around the y-axis while maintaining stability and avoiding obstacles.\label{fig:quadrotor-visual-result}}
\end{figure}

We tested the proposed method for optimizing a safe trajectory for this quadrotor dynamics. The task (see Fig.~\ref{fig:quadrotor-ex}) involves moving from the initial position $(3, 0.25, 2)$ to the position $(2, 5, 2)$ within 4 seconds, with a fixed time step of $\Delta t = 0.02$. Two rectangular boxes with dimensions $(1.5, 0.1, 3.0)$ are located at $(0, 2.5, 3.0)$ and $(3.6, 2.5, 3.0)$, which must be avoided. In addition to obstacle avoidance, we included a configuration attainment constraint to test for gimbal-lock and forced the robot to achieve $R_y(90)$ while passing between the boxes. Additionally, due to the nature of the actuators, rotor speeds are bounded between 0 and maximum rpm. 

Figure \ref{fig:quadrotor-visual-result} shows the resultant trajectories using our proposed method. The robot successfully achieves the $R_y(90)$ rotation ($\theta$ angle) without experiencing gimbal-lock between time steps 1.2 - 1.9 sec, during which it required zero input. Although the robot's initial and goal z-positions are the same, it increases its z-position at the beginning of the trajectory to prevent falling due to the period of zero input (Figure \ref{fig:quadrotor-visual-result}).

\section{Conclusion}
\label{sec:conclusion}
Optimization of control policies on matrix Lie groups presents a significant challenge in control and robotics, while offering numerous practical applications. 
In this work, we have proposed a novel approach for tackling this problem using an augmented Lagrangian based constrained discrete \ac{ddp} algorithm. 
Our method involves lifting the optimization problem to the Lie algebra in the backward pass, facilitating the computation of gradients for objective and constraint functions within the corresponding tangent space. 
In the forward pass, we retract back to the manifold by integrating the dynamics using the optimal policy obtained in the backward pass.

A key contribution of this work is the development of a general approach for handling nonlinear constraints within the \ac{ddp} framework across a wide range of matrix Lie groups.
The method thus surpasses previous \ac{ddp} methods, which were constrained to specific matrix Lie groups, by providing a more versatile and widely applicable solution.
Additionally, compared to existing direct optimization methods, our approach demonstrates increased computational efficiency, achieving faster convergence rates with fewer iterations due to its inherent incorporation of the dynamics and leveraging second-order information for more accurate updates.
We also demonstrated the method's effectiveness in handling external disturbances through its application as a Lie-algebraic feedback control policy on SE(3). 
The results show that the approach is able to effectively handle constraints and maintain its stability in the presence of external disturbances.

One of the limitations of our work is the assumption that the deterministic transition dynamics of the controlled systems are assumed to be known. 
Future research could address this by including the learning of these dynamics within the matrix Lie group representation. 
Additionally, incorporating uncertainties from both the dynamic models and measurements into the planning, and investigating closed-loop uncertainty propagation similar to the approach used for Euclidean models in \cite{Alcan2022} could significantly enhance the robustness of the proposed approach. 
These advancements would further enable the method to perform more effectively in complex, dynamic and unknown environments.


\end{document}